\documentclass[journal=jpclcd,manuscript=letter,layout=onecolumn]{achemso}

\usepackage[dvipsnames]{xcolor}
\usepackage{bm}
\usepackage{color} 
\usepackage{dcolumn} 
\usepackage{graphicx} 
\usepackage{epstopdf}
\usepackage{hyperref} 
\usepackage{amsmath}
\usepackage{pslatex}
\usepackage{siunitx}
\usepackage{subfigure} 

\def\PoP{\cal $\mathcal{P}o\mathcal{P}$}
\def\2ww{$2\omega$-$\omega$}

\author{B. Krebs}
\affiliation{Institute of Physics, University of Rostock, 18059
  Rostock, Germany}
  
\author{V. A. Tulsky}
\affiliation{Institute of Physics, University of Rostock, 18059
  Rostock, Germany}    

\author{L. Kazak}
\affiliation{Institute of Physics, University of Rostock, 18059
  Rostock, Germany}  
  \altaffiliation{Present address: Institute for Quantum Optics, Ulm University, 89081 Ulm, Germany}
  
\author{M. Zabel}
\affiliation{Institute of Physics, University of Rostock, 18059
  Rostock, Germany}
  
\author{D. Bauer}
\affiliation{Institute of Physics, University of Rostock, 18059
  Rostock, Germany}

\author{J. Tiggesb\"aumker}
\email{josef.tiggesbaeumker@uni-rostock.de}
\affiliation{Institute of Physics, University of Rostock, 18059
  Rostock, Germany}
\affiliation{Department ``Life, Light and Matter'', University of Rostock, 18059
  Rostock, Germany }

\title{Phase-of-the-Phase Electron Momentum Spectroscopy on Single Metal Atoms in Helium Nanodroplets}
  
\begin{document}

\begin{tocentry}
    \includegraphics{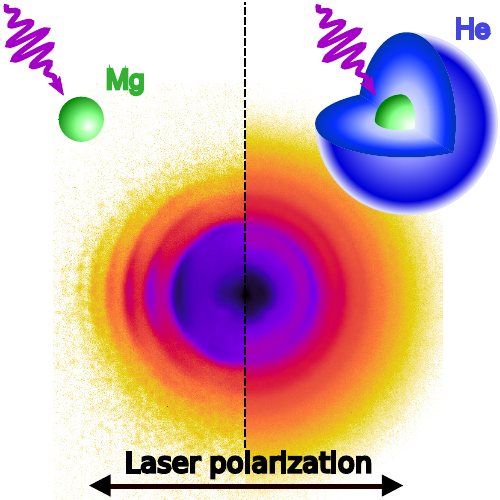} 
\end{tocentry} 

\begin{abstract}

Magnesium atoms fully embedded in helium nanodroplets are exposed to two-color laser pulses, which trigger multiphoton above-threshold ionization (ATI). This allows to exemplarily study the contribution of a dense, neutral and finite medium on single electron propagation. The angular-resolved photoelectron spectra show striking differences with respect to results obtained on free atoms. Scattering of the individual Mg photoelectrons, when traversing the neutral helium environment, causes the angular distribution to become almost isotropic. Furthermore, the appearance of higher energy electrons is observed, pointing out the impact of the droplet on the concerted emission process. Phase-of-the-phase spectroscopy, however, reveals a marked loss in the \2ww phase dependence of the electron signal. Taking into account sideband formation on a quantitative level, a Monte-Carlo simulation which includes laser-assisted electron scattering can reproduce the experimental spectra and give insights into the strong-field-induced electron emission from disordered systems. 

\end{abstract}

In strong-field ionization, the interaction of the liberated electron with the remaining ion has drawn considerable attention~\cite{GalARPC12,AmiRPP19,LinNRP19}. In particular, the intensity regime differentiating multiphoton and field-driven ionization\cite{KelSPJETP65} is of general interest. The analysis of electron emission from strong field interactions provides detailed information on the ionization dynamics, which allow to probe matter on the shortest spatial and temporal scales.\cite{RamARPC16} For example, the different birth times of the electron wave packets lead to interferences, which can be analyzed by recording the photoelectron angular distribution (PAD).\cite{HePRL11} Outstanding electron features are carpet structures\cite{KorPRL12} and holographic side lobes.\cite{HuiS11} Further, the PAD provides details on the ultrafast electron dynamics even at lower laser intensities. The appearance of above-threshold ionization (ATI) patterns modified by Freeman resonances may serve as a prominent example in the multiphoton regime.\cite{FrePRL87}

In order to fully resolve the electron dynamics experimentally, the driving laser field has to be adjusted on a time scale shorter than an optical cycle. Experiments using two-color pulses are appealing, since the superimposed laser fields allows for a control of the relative intensity and phase as well as polarization~\cite{KimPRL05,XiePRL17,TulJPB20}. For resolving phase dependencies covering the entire PAD, phase-of-the-phase  electron momentum spectroscopy (\PoP) has been introduced recently~\cite{SkrPRL15}. \PoP\ is a powerful tool  to identify and analyze coherent dynamics, as masking incoherent contributions to the spectra are effectively suppressed. Phase-dependent changes in the PAD provide a sensitive marker for interfering influences on the electron trajectories\cite{FigRPP20}. Recently, a characteristic checkerboard pattern in the ATI spectrum of xenon was identified using \PoP\ and analytically described within the strong-field approximation.\cite{AlmJPB17} In order to resolve fine details of the electron dynamics, \PoP\ utilizes the precise control of the relative phase between the two-color components and the differentiation into phase-dependent and phase-independent contributions to the  signals. The power of the \PoP\ has been demonstrated in strong-field ionization of atoms and molecules.\cite{SkrPRL15,AlmJPB17,WuePRA17,TulPRA18,TulJPB20} Even relativistic two-color-field-induced pair production in vacuum has been discussed on a theoretical level~\cite{BraPRA20}.

\begin{figure}
	\resizebox{8.6cm}{!}{\includegraphics{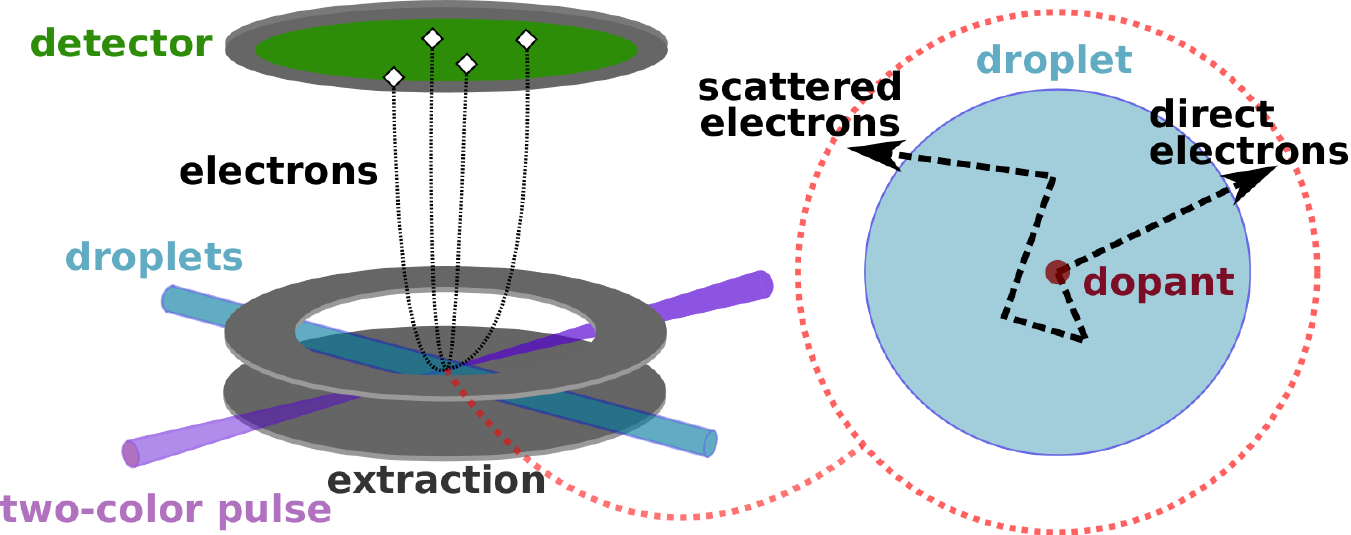}}
	\caption{Schematic view of the experimental setup to study phase-dependencies in the photoemission from single metal atoms embedded in helium nanodroplets triggered by strong linearly polarized two-color laser fields. Momentum-resolved electron signals are obtained by velocity map imaging (VMI). The expanded view on the ionization dynamics inside the droplets highlights photoemission pathways taken into account in the computational analysis.}
    \label{fig:KreArXive22-f1}
\end{figure}

In the present study, the \PoP\ method is applied to the strong-field-induced photoemission from atomic impurities within a medium of bulk density, see the scheme depicted in Fig.~\ref{fig:KreArXive22-f1}. More precisely, multiphoton ionization from single atoms enclosed by a size-limited dense medium is investigated, wherein single electrons propagate through the neutral environment while under the influence of the strong laser field. So far, such a phase-sensitive ionization scenario has not been experimentally considered. As surrounding medium, we choose helium nanodroplets, which offer unique low-temperature conditions for spectroscopy~\cite{GreS98}. 

We refer to~\cite{MudIRPC14, ZieIRPC15, VerAdvP19} for recent reviews in the field. The strong-field response of helium nanodroplets has been studied in context of the nanoplasma response and gives insights into, e.g., collective effects~\cite{TigPCCP07, KriPCCP14} and relaxation dynamics~\cite{KelJCP19, KelPRL20}. In the present investigation, we inspect the multiphoton intensity regime and focus on the signatures of ATI in the ionization of magnesium atoms embedded in helium nanodroplets. ATI originates from electron waves periodically emitted at different cycles of the laser pulse. Only events separated by an integer number of the optical cycle interfere constructively. The corresponding pattern has been found in atoms and molecules and shows up as regularly spaced features separated by the laser photon energy~\cite{EbePhysRep91}. In the following, we will show that in comparison to free atoms, the two-color photoelectron studies from embedded atoms  reveal  severe  changes  in  the angular pattern and an appearance of higher energy peaks which point at laser-assisted electron-helium scattering. Further, these findings are accompanied by a complete loss in phase-dependencies, as \PoP\ reveals.

The experiment used to investigate the strong-field electron dynamics of atoms embedded in helium nanodroplets is depicted in Fig.~\ref{fig:KreArXive22-f1}. Briefly, a molecular beam of droplets is produced by supersonic expansion of pre-cooled helium gas (\SI{10.5}{\kelvin}) through a \SI{5}{\micro\meter} nozzle into vacuum at a backing pressure of \SI{20}{\bar}, which corresponds to a mean droplet size of $ N_{\mathrm{avg}}=54\,000$. On its way to the interaction region the molecular beam passes a resistively heated oven filled with magnesium. The average number of Mg impurities per droplet is about 0.1. This establishes single-atom doping conditions and ensures that contributions from dimer formation are negligible. In addition, a mechanical shutter is installed between the droplet source and the pick-up region allowing to block the droplet beam and conduct measurements on free Mg atoms, while all other conditions remain unchanged. After the pick-up chamber, the droplet beam enters the velocity-map-imaging spectrometer~\cite{SkrIJMS14} where it intersects with the laser beam that is polarized parallel to the detector plane. The laser system provides 110\,fs-long linearly polarized pulses at central wavelength of 800\,nm. Further, the second harmonic 2$\omega$ is produced in a BBO-II crystal and propagates co-linearly with the fundamental. The temporal overlap between 2$\omega$ and $\omega$ pulses is controlled by an array of calcite plates, leading to the formation of two-color pulses. Focusing onto the target is achieved through a $f=\SI{250}{\milli\meter}$ metallic mirror. In contrast to commonly used setups, the 2$\omega$ component serves to ionize the target, whereas $\omega$ modifies the pulse shape. 
The intensities of the harmonics are set equal to $I_{2\omega} = 5.6\cdot10^{13}$\,W\,cm$^{-2}$ and $I_\omega = 0.03\cdot I_{2\omega}$, respectively, and correspond to the Mg multiphoton ionization regime. These intensities refer to a Keldysh parameter~\cite{KelSPJETP65} of $\gamma=2.1$ for the dominant 2$\omega$ component. Such laser conditions and the large difference in ionization potentials~\cite{NIST} of Mg (7.65 eV) and He (24.59 eV) allow above-threshold ionization of Mg without dopant-induced nanoplasma formation~\cite{MikPRL09, HeiNJP16}.
The relative phase shift $\varphi$ between the \2ww laser components, responsible for the asymmetry of the laser field, is controlled with sub-cycle precision by a pair of movable fused silica wedges installed in front of the experimental chamber. We recorded $200$ PADs at various values of $\varphi$, equally distributed between $0$ and $6\pi$.

In order to obtain the \PoP\ spectra, the PADs are Fourier-transformed with respect to $\varphi$. Due to the $I_{\omega}/I_{2\omega}$ ratio being small, this Fourier decomposition can be well represented by the two leading terms:
\begin{equation}
    Y (\textbf{p} ,\varphi) \simeq  Y_0 (\textbf{p}) + Y_1 (\textbf{p})\cos[\varphi + \Phi_1(\textbf{p})].
    \label{eq:FT}
\end{equation}
Here, $Y_0(\textbf{p})$ represents the phase-averaged momentum spectrum (PA) and bundles the information about the processes taking place in the two-color laser field disregarding possible phase-dependencies. Conceptionally, the result compares to few-cycle experiments, which are conducted without phase stabilization or phase tracing~\cite{SueRSI11}. The second term in Eq.~(\ref{eq:FT}) includes $Y_1(\textbf{p})$ and $\varPhi _1(\textbf{p})$, which equal the relative-phase-contrast (RPC) and phase-of-the-phase (PP) spectra, respectively. RPC and PP provide momentum-resolved information about the degree of phase dependencies encoded in distinct features (see, e.g., ~\cite{SkrPRL15} for more details).

\begin{figure}[t]
    \centering
	\resizebox{0.9\columnwidth}{!}{\includegraphics{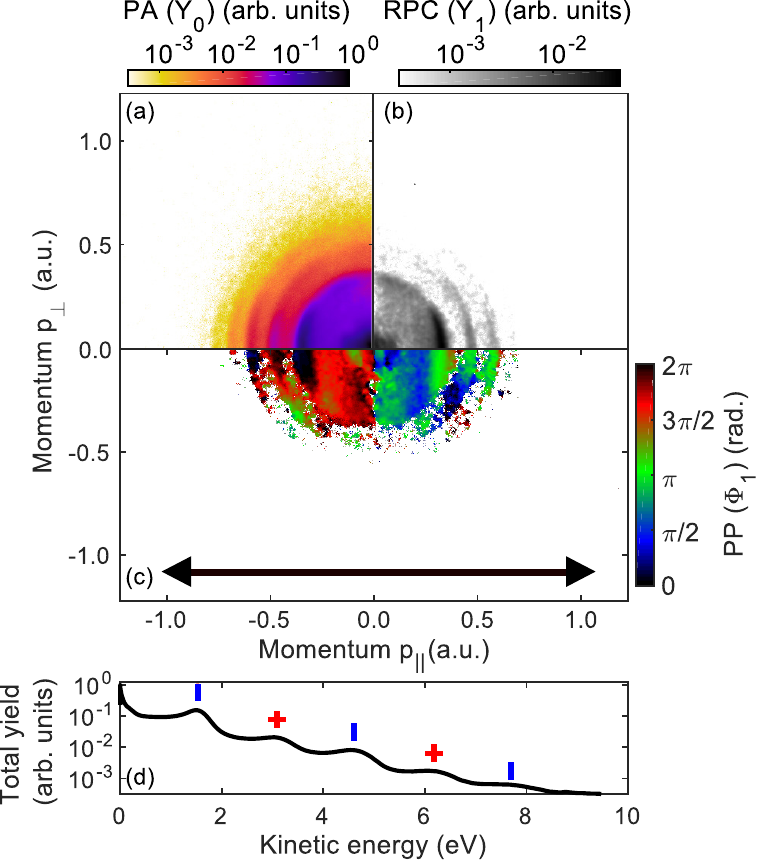}}
	\caption{(color online) \PoP\ spectra (a: PA ($Y_0$), b: RPC ($Y_1$), c: PP ($\Phi_1$) of atomic magnesium extracted from the photoelectron spectra obtained by phase-sensitive two-color ionization (\,$\hbar\omega=1.55$\,eV, $I_{2\omega}=5.6\cdot10^{13}$\,W\,cm$^{-2}$, $I_\omega = 0.03 \cdot I_{2\omega}$). Due to symmetry reasons, it is sufficient to display only quarter segments of the spectra. The PA and RPC signals are normalized to the maximum of $Y_0$. PP values are not shown, whenever the corresponding RPC result drops below $10^{-4}$. The arrow indicates the polarization axis of the \2ww\ laser field. (d) Angle-integrated PA yield as function of electron energy, which is characterized by ATI peaks. Even and odd ATI orders are emphasized by (+) and ($|$), respectively. Note the logarithmic scaling in the yields.}
    \label{fig:KreArXive22-f2}
\end{figure}

The decomposition process (Eq.~\ref{eq:FT}) is demonstrated by measuring the \PoP\ spectra of free magnesium atoms, see Fig.~\ref{fig:KreArXive22-f2}. The PA ($Y_0$) shows a clear ATI pattern, which is strongest along the laser polarization axis as expected. Although the photon energy of the stronger laser component is $2\,\hbar\omega=3.10$\,eV, the ATI features are spaced by $\hbar\omega = 1.55$\,eV, which corresponds to the photon energy of the weaker component. The odd rings, however, are relatively weaker than the even ones, being enforced by the $2\omega$ field (see Fig.~\ref{fig:KreArXive22-f2} (d)).

Inspecting the phase-of-the-phase reveals that the PP signals for positive and negative momenta $p_{\parallel}$ differ by a phase shift of $\pi$. This characteristic feature obtained by  {\PoP} can be explained from a semi-classical point of view: the strong $2\omega$ field is responsible for the instantaneous ionization probability and the weak  $\omega$ streaking field defines the asymmetry of the yield being opposite for phase values of $\varphi$ which differ by $\pi$.\cite{SkrPRL15}. 
For a given half-space of momenta, e.g. $p_{\parallel}>0$, additional changes in the PP may reflect the interplay between different ionization pathways.\cite{AlmJPB17}
From this finding one can deduce that the information imprinted on the launched electron wave packet by the asymmetrical laser field is preserved until detection. The clearly discernible signals therefore reflect an almost full coherence process, which can be linked to the \2ww\ phase shift. A deeper analysis of the magnesium two-color ionization in the multiphoton regime is a task on its own and goes beyond the scope of the present letter. For the further treatment, however, it is relevant that the multiphoton ionization process under study exhibits a strong dependence on $\varphi$, which is expressed in clear RPC and PP signals.

\begin{figure}[t]
    \centering
	\resizebox{0.9\columnwidth}{!}{\includegraphics{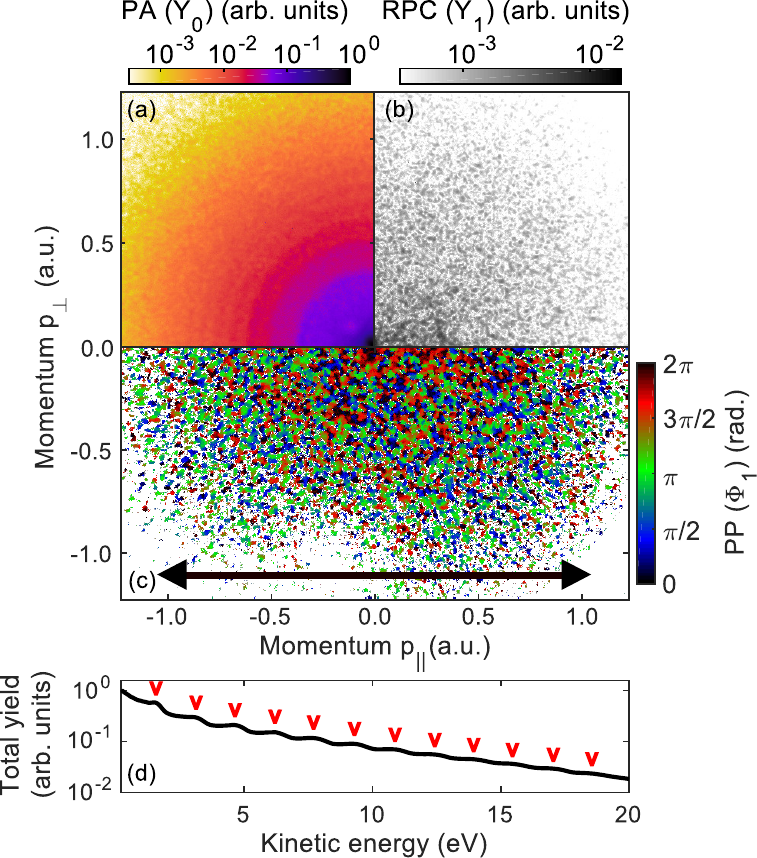}}
	\caption{
	\PoP\ spectra of single magnesium atoms embedded in helium droplets, obtained under laser conditions as indicated in Fig.~\ref{fig:KreArXive22-f2}. The PA spectrum (a) shows an almost isotropic electron emission. Compared to free Mg (Fig.~\ref{fig:KreArXive22-f2}) we note the appearance of higher energy peaks and a complete loss in the phase-dependencies between \2ww\ laser field and electron momentum. The latter manifests as a non-structured granular phase signal in the RPC and PP spectra.}
    \label{fig:KreArXive22-f3}
\end{figure}

Based on these studies, experiments on magnesium embedded in helium droplets were carried out under identical laser conditions. The results presented in Fig.~\ref{fig:KreArXive22-f3} can be summarized as follows:

\begin{itemize}
    \item[(i)] The angular distribution becomes nearly isotropic (cf. Figs.~\ref{fig:KreArXive22-f2}\,(a) and~\ref{fig:KreArXive22-f3}\,(a));
    \item[(ii)] A careful check of the experimental spectra reveals no evidence that phase-dependent features are still present (cf.  Figs.~\ref{fig:KreArXive22-f2}\,(b,c) and~\ref{fig:KreArXive22-f3}\,(b,c));
    \item[(iii)] The spectrum extends to higher energy. Up to 12 photoelectron peaks can now be resolved (cf. Figs.~\ref{fig:KreArXive22-f2}\,(d) and~\ref{fig:KreArXive22-f3}\,(d));
    \item[(iv)] No pronounced energy shifts of the ATI peaks are obtained.
\end{itemize}

With regard to the PA, there are indications in the PP and RPC spectra on signals of larger momentum.  However, in comparison to free Mg, the RPC to PA ratio in droplets diminishes by at least one order of magnitude, which manifests in non-structured granular distribution of events in the RPC and PP spectra obtained from embedded Mg. Even the characteristic PP shift of $\pi$ is not present. One can draw the conclusion that all features of phase-dependencies have disappeared and thus the previously observed coherence of the process has been completely lost. It is, however, rather unlikely that the \2ww\ laser field no longer has an impact on the ionization pathways in embedded Mg. 

The increase in the electron energy and the change in the angular distribution shows similarities to recently reported strong-field electron dynamics studies on dielectric spheres\cite{ZheNPhys11, LiuACSPhot20} and atomic clusters.\cite{SchSR16, PasNCom17} In dielectric nanospheres, dynamical near-field enhanced backscattering of electrons after tunneling at the surface leads to directed electron acceleration.\cite{ZheNPhys11, LiuNJP19} The scattering potential, however, stems from charge separation of ions and electrons at the interface. Such behavior is rather unlikely due to the low polarizability of helium.\cite{ToeAC04} In clusters being subject to higher laser intensities ($\gamma\le1$), the avalanche-type charging process{~\cite{DoePRL10}} results in a rapid formation of a nanoplasma where the dynamics is dominated by the collective response of the quasi-free electrons. In this scenario laser-assisted electron acceleration,\cite{FenPRL07a, SaaPRL08} including rescattering scenarios,\cite{SchSR16, WanPRL20} are in general only effective at high nanoplasma temperatures. Moreover, phase dependencies in the electron signals are expected.\cite{PasNCom17,WanPRL20} Regarding experiments on helium nanodroplets, enhancements in the ATI emission in extreme ultraviolet (XUV) pump near-infrared probe experiments have been reported recently.\cite{MicPRL21} The origin, however, was attributed to the collective electron dynamics triggered by the initial resonant XUV photoexcitation of the pure droplets. Hence, in the above experimental scenarios, cooperative and collective effects dominate the entire electron dynamics. Since such excitations in the intensity regime under inspection are unlikely to occur, we conclude that the observed features arise from the propagation of single active electrons through a neutral helium environment.

For further processing of the data, the target conditions have to be clarified. It is known that Mg atoms reside near the center of the  droplet~\cite{HerPRB08}. Hence, the dopant is enclosed by a spherical shell of helium atoms. Further, underpinned by the considerable difference in the ionization potentials between Mg and He and the lack of low-energy helium excitation levels, electron emission solely originates from the dopant. Observing no photoemission from pure droplets under the chosen laser conditions proves this assertion. These special properties suggest to divide the ionization dynamics in droplets into independent processes that is multiphoton ionization from the impurity atoms and scattering of the electrons when transversing the enclosing helium shell.

With respect to the strong field electron dynamics in dense, neutral and disordered matter, the theoretical modeling is not yet entirely convincing. Electron motion in liquid noble gases\cite{BoyJCP15,BoJPD16,WhiPSST18} and the mobility of positrons in liquid helium~\cite{CocJPB20} have been simulated.  All studies report on significantly modified scattering cross sections compared to the gas phase. To the best of our knowledge, however, related studies on femtosecond-resolved electron motion in liquid helium, which includes e-He scattering have not been conducted so far. Therefore, we opted to fall back on data recorded by gas phase scattering experiments.\cite{JanJPB76} Further, the presence of a laser field can lead to changes in the scattering cross sections~\cite{ChoPLA75}, usually  described on the basis of Kroll-Watson theory.\cite{KroPRA73} Under conditions, where the electron energies are significantly higher than the photon energy, good agreements are obtained.\cite{KanAtoms19} Note that Kroll Watson theory has recently been used under similar conditions\cite{TreNCom21} and has shown good agreement, although the applicability of this theory to such scenarios has still to be proven. We thus adopt the current knowledge and analyzed the change in the distribution of photoelectrons emitted from embedded Mg within a Monte-Carlo (MC) simulation. In addition, laser-assisted electron scattering (LAES) is included,\cite{BunJETP65, KroPRA73} see, e.g., ~\cite{KanAtoms19, MasRPP93, EhlPR98} for reviews on the topic. A detailed description of the MC simulation procedure can be found in the supplementary material at URL.

\begin{figure}[t]
    \centering
	\resizebox{0.99\columnwidth}{!}{\includegraphics{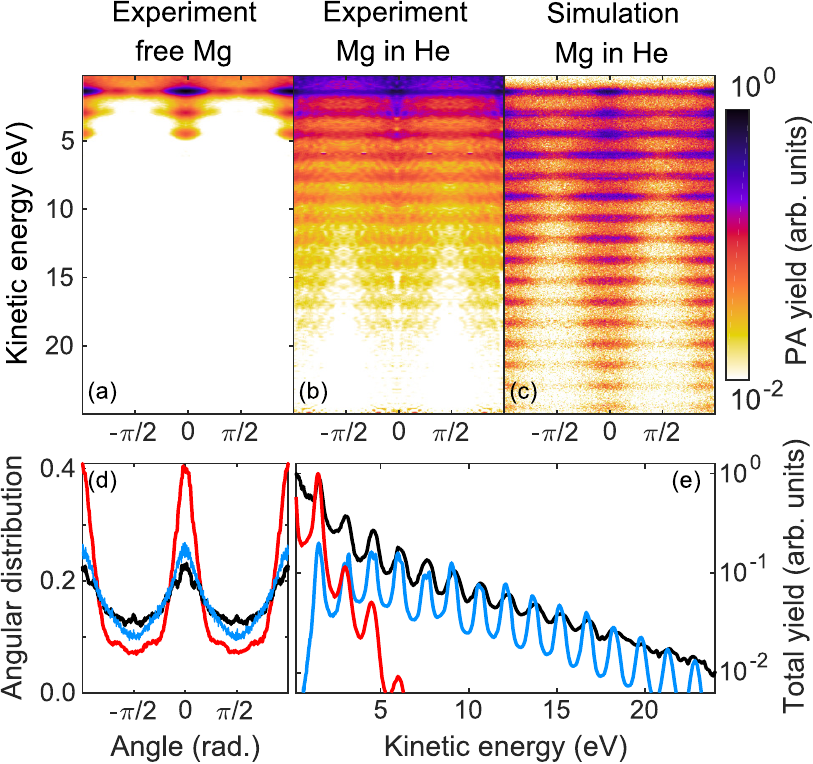}}
	\caption{Inverse Abel transformed photoelectron spectra of free Mg (a) and embedded Mg (b) plotted vs energy and angle. (c) Result of a Monte-Carlo simulation as described in the text. Corresponding angular distributions (d) and energy spectra (e): free Mg, red; embedded Mg, black; MC simulation, blue (vertically adjusted to highlight the agreement between the slopes).}
    \label{fig:KreArXive22-f4}
\end{figure}

The results of the corresponding simulation are presented in Fig.~\ref{fig:KreArXive22-f4}\,(c-e) and are compared to the VMI measurements being subject to an inverse Abel transform using the onion peeling procedure~\cite{RobRSI09}. Based on the strong anisotropy obtained for Mg (Fig.~\ref{fig:KreArXive22-f4}\,(a)), e-He scattering turns out to be the crucial factor determining the final emission direction. In the simulations we found that 28\% of the electrons undergo one scattering event, while 58\% scatter multiple times. Only about 14\% of electrons leave the droplets without any further interaction (see supplementary material). The latter would result in a small number of electrons carrying phase information, whose detection is below our current experimental sensitivity. Further, the simulated yield from embedded Mg appears to be almost isotropic and deviates from the experiment only at momenta parallel to the laser field polarization, see black line in Fig.~\ref{fig:KreArXive22-f4}\,(e). Note, that inevitably the inverse Abel transform of the experimental data leads to an artificial enhancement of the signals along this axis.

The formation of the high-energy peaks is reproduced by including the influence of the laser field on the scattering process~\cite{KroPRA73, ChoPLA75}, see Fig.~\ref{fig:KreArXive22-f4}(e). The exponential slope of the simulated energy distribution matches the experimental result apart from the low energy region. Considering various origins for this deviation, we came to the following conclusions. Since the density of free Mg in the interaction region is significantly lower compared to the target density provided by the helium droplet beam a contribution of free Mg to the signals can be excluded. A misjudgment with respect to the droplet size distribution can also be ruled out. Signals from unscattered electrons should preferably arise from emission from small droplets, i.e., when the scattering mean free path is comparable to the droplet size. However, the probability that these droplets will contain an Mg atom is also negligible due to the small capture cross-section. (see the detailed explanation in the supplementary material at URL). Thus, the enhancement around the first peak can be attributed to signals from residual gas expected to show up at electron energies below 4\,eV.

The appearance of electrons with higher energies suggests for an energy exchange between helium environment and laser field. One might expect that the ATI progression changes if excited states of helium are involved or if direct ionization can take place. However, no Freeman resonances~\cite{FrePRL87} or a marked change in the slope above $19.8$\,eV being the first excited state of helium~\cite{NIST} are observed, respectively. Hence, LAES at neutral He atoms is responsible for the increase of electron energies. This leads to the appearance of sidebands to the photoelectron peaks, separated by the photon energy.\cite{KanAtoms19} This effect is well reproduced by the simulations, see Fig.~\ref{fig:KreArXive22-f4}. Interesting aspects to be included in advanced simulations are multi-center scattering and medium-affected ionization as a result of the bulk helium density conditions. Taking the obvious limitations of the current approach into account, the agreement between experiment and modeling is remarkable and motivates to improve the model in terms of a comprehensive description of the dynamics.

Laser-assisted electron scattering in helium nanodroplets, see also Ref.\cite{TreNCom21} bridges the gap between similar processes in gases{~\cite{CerJPB06, WilNJP14}} and single molecules.\cite{FetPRA19} In experiments on dense vapors, single and multiple LAES events have been observed. However, the scattering probability and subsequent electron energy gain in droplets is enhanced due to the bulk density conditions. This allows for the emission of higher energy electrons at relatively low laser intensities. In molecular physics, laser ionization from diatomic molecular ions, e.g., H$_2^+$,  with subsequent electron re-scattering at one or the other atom in the parent molecule, has been discussed.\cite{FetPRA19} Compared to single  atoms, an increase in the re-scattering energy cutoff at large interatomic distances was found, whereas atomic-like coherent scattering takes place at equilibrium distances. Interestingly, the latter  resembles the conditions in the droplets with an interatomic spacing of only a few Angstroms. Still, LAES takes place rather than coherent scattering, as indicated by the lack of corresponding signatures in the \PoP\ spectra.

In conclusion, \PoP\ gives insight into the strong-field ionization dynamics in a dense environment. In particular, we studied the above-threshold ionization of isolated Mg and Mg atoms embedded in He nanodroplets. The obtained spectra, which result from single electrons, substantially differ when propagating in the neutral medium. The inspection of the angular-resolved electron momentum spectra and accompanying Monte-Carlo simulations reveal that the electron trajectories are modified by multiple laser-assisted scattering with the helium environment, which enhances the isotropy and leads to higher energy sidebands. Moreover, the \PoP~ analysis reveals a complete loss in the {\2ww} phase dependence, indicating that the increase of the photoelectron energy predominantly stems from incoherent LAES on helium atoms, rather than coherent re-scattering on the residual Mg parent ion. The comprehensive relevance of such incoherent processes in the context of subsequent scattering events has been pointed out.{~\cite{CerLP09}} 
The considerable increase in the scattering mean free path in strong field photoemission through LAES will in future allow to have an even more detailed view on the ionization dynamics of systems enclosed by disordered media. Both scattered and unscattered electrons contribute to the \PoP-signals, but carry complementary information,\cite{TulJPB20} i.e., inner atomic dynamics and medium interaction. 
By using smaller droplets, the unscattered fraction will even increase, which provides a handle to resolve the underlying processes. It should be pointed out that the scenario under consideration can also be realized in core-shell systems, e.g., O$_2$Ar$_N$Ne$_M$.\cite{LaaJCP08} This results in a variety of possibilities to study the ionization dynamics of embedded atoms, molecules and clusters. Special to helium, unique ultracold conditions can be provided.\cite{GreS98} For example, the ionization dynamics of chiral molecules\cite{ComJPCL16, BeaSci17} within a superfluid environment with respect to circular dichroism can be studied by being subject to circular polarized two-color laser fields. 

With these experiments, we tackle the challenge to bridge the gap  between the rather well understood  strong-field laser physics of free atoms in vacuum and the much more complex physics of photoelectron emission from atoms embedded in a condensed, nanometer-size environment, in that way paving the way towards atto-nano science~\cite{Hommelhoff2015, CiaRPP17}.

The Deutsche Forschungsgemeinschaft (BA 2190/\,10, TI 210/\,7, TI 210/\,8) is gratefully acknowledged for financial support.

\providecommand{\latin}[1]{#1}
\makeatletter
\providecommand{\doi}
  {\begingroup\let\do\@makeother\dospecials
  \catcode`\{=1 \catcode`\}=2 \doi@aux}
\providecommand{\doi@aux}[1]{\endgroup\texttt{#1}}
\makeatother
\providecommand*\mcitethebibliography{\thebibliography}
\csname @ifundefined\endcsname{endmcitethebibliography}
  {\let\endmcitethebibliography\endthebibliography}{}

\end{document}


\section{Monte-Carlo simulation}

The procedure to conduct the Monte-Carlo calculation is described in~\cite{TulJPB20} and adapted to the present scenario. With respect to the qualitative nature of the model and due to the azimuthal symmetry of the photoelectron distribution as a result of the linearly polarized laser field, the simulations are restricted to only two dimensions.

\subsection{Initial condition}
According to density functional theory~\cite{HerPRB08}, Mg resides at the center of the droplet, which in the following defines the origin of each ionization event  ($N_{\mathrm{e}} = 10^{6}$ electrons in total) in the simulation. The initial velocity is chosen according to the distribution recorded for free Mg atoms in the considered laser field which we obtain from data shown in Fig.~2\,(a) in the main text by performing the deconvolution via polar onion peeling~\cite{RobRSI09}. This procedure gives the spectrum of electrons within the plane parallel to the laser polarization (shown in Fig.~4(a) in the main text) and validates the 3D $\to$ 2D reduction of our model. In order to suppress signals originating from residual gas in the interaction region, we restrict our analysis to electron energies  $E=p^2/2m>0.2$ eV ($\vert {\bf p}\vert >0.12$ atomic units).

The droplet size is chosen individually for each electron according to a log-normal droplet size distribution~\cite{HarPRB98} 
\begin{equation}
    P_1(N_{\mathrm{He}}) = \frac{1}{\sqrt{2\pi}}\frac{1}{N_{\mathrm{He}}\,\delta}\exp\left[-\frac{(\ln N_{\mathrm{He}}-\mu)^2}{2\delta^2}\right]
\end{equation}
with parameters $\mu=\ln N_{\mathrm{avg}}-\delta^2/\,2$, $\delta$=0.626 (corresponding to an average number of $N_{\mathrm{avg}}$=54\,000 helium atoms per droplet, which corresponds to a mean radius $8.18$\,nm for empty droplets). The number of dopand atoms being captured per a droplet of mean size is $\lambda=0.08$ and in general is known to obey the Poisson distribution~\cite{LewJCP95, BueEPJD11}, thus, the final distribution of number of atoms $N_{\mathrm{He}}$  in a helium droplet that embeds a single Mg atom reads
\begin{equation}
P(N_{\mathrm{He}}) = P_1(N_{\mathrm{He}})\cdot \frac{\lambda}{1!} e^{-\lambda}
\end{equation}
where the Poisson distribution parameter $\lambda$ reads
\begin{eqnarray}
\lambda &=& \pi R^2 L n_{\mathrm{Mg}}\\ 
R &=& [3N_{\mathrm{He}}/(4\pi n_{\mathrm{He}})]^{1/3} = 2.22\,  N_{\mathrm{He}}^{1/3}\,[\mathrm{\AA}].
\end{eqnarray}
Basically, it reflects the number of Mg atoms that are in the geometrical volume that a droplet of size $N_{\mathrm{He}}$ covers in the pickup chamber. In our case, $L=3.35$\,cm and $n_{\mathrm{Mg}} = 1.072 \cdot 10^{16}$\,cm$^{-3}$. We note, however, that the influence of the droplet size distribution on the final results had only a minor effect.

\subsection{Random walk in a droplet}
Subsequently, the random walk of each electron within the droplet is treated by calculating the probability of scattering after traveling a certain spatial distance $s$ as 
\begin{equation}
    w(s,E) = 1 - \exp(-n_{\mathrm{He}} \sigma(E) s)
\end{equation}
where $\sigma(E)$ denotes the total scattering cross section. The concentration of helium $n_{\mathrm{He}}$ is assumed to be homogeneous and equal to the bulk value of $n_{\mathrm{He}}=2.18\cdot10^{28}$\,cm$^{-3}$~\cite{BarJLTP06}. The spatial distance between scattering events
\begin{equation}
    s = -\frac{1}{n_{\mathrm{He}}\sigma(E)}\ln X
\end{equation}
is obtained by choosing a random number $X$ uniformly distributed in the range $[0,1)$. If $s$ is smaller than the current distance to the surface of the droplet, scattering takes place. The scattering angle $\theta$ is then chosen according to the differential cross section $d\sigma(\theta,\epsilon)/d\Omega$.

\begin{figure*}
    \centering
	\includegraphics[width=0.66\linewidth]{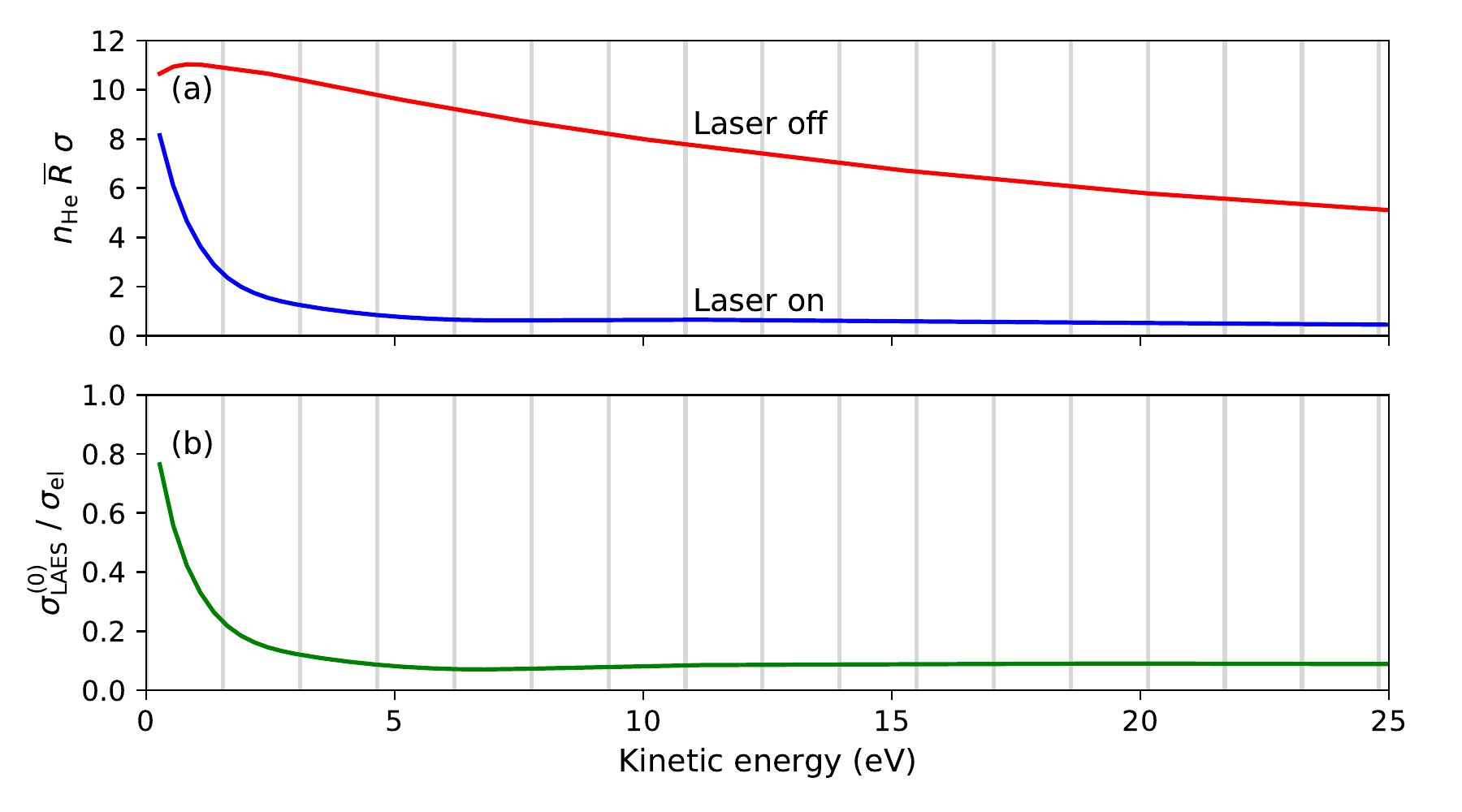}
	\includegraphics[width=0.33\linewidth]{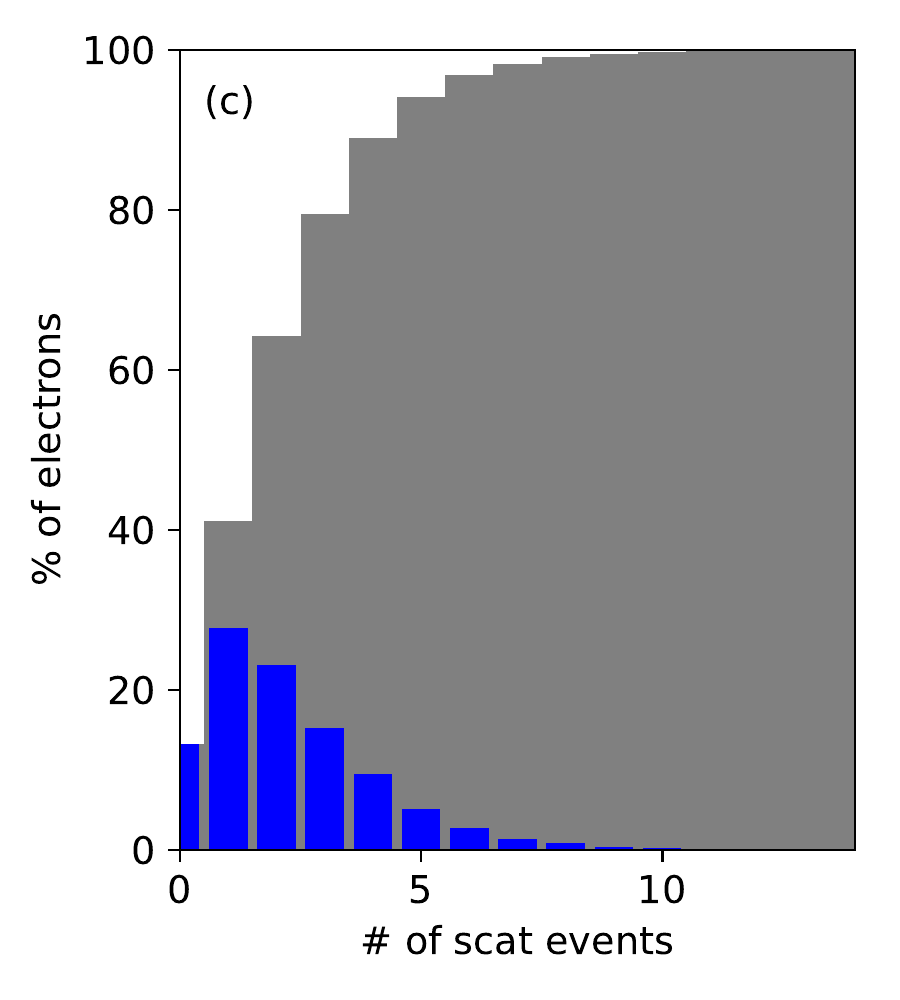}
	\caption{(a) Characteristic dimensionless scattering parameter being a product of helium atoms concentration, $n_{\mathrm{He}}= 2.18\cdot 10^{28}$cm$^{-3}$, nanodroplet mean radius, $R= 8.4$ nm ($N= 54\,000$), and the total cross section $\sigma$, for scattering of an electron on a neutral helium atom with (blue line) and without (red line) the presence of a strong laser field. A channel with $\nu=0$ photon emission/absorption is considered. (b) Ratio of the LAES cross sections to the cross sections under field-free conditions. Laser parameters: $\omega = 1.55$ eV, $\omega_2 = 2\omega$, $I_{2\omega}= 5.6\cdot 10^{13}$W/cm$^2$, $I_{\omega}= 0.03\cdot I_{2\omega}$, co-linear polarizations. (c) Percentage of electrons which scatter a certain number of times (blue histogram) and percentage of the electrons which have experienced up to a certain number of LAES scattering events (grey histogram).}
    \label{fig:KreJPCL21-fS1}
\end{figure*}

In extending the model of~\cite{TulJPB20},  laser-assisted electron-helium scattering is treated by using the Kroll-Watson formula~\cite{KroPRA73} generalized for two-color laser fields. The off-shell energy $\epsilon(\theta,E)$, at which the laser-free differential cross section $d\sigma(\theta, \epsilon)/d\Omega$ is used in that theory, depends on the direction of scattering, and the number of photons involved in a process, thus governing the formation of the sidebands. The total cross sections $\sigma(E)$ are replaced with those computed according to the Kroll-Watson formula for $\nu=0$ photon absorption (see Fig.~\ref{fig:KreJPCL21-fS1}). Field-free values are taken from~\cite{JanJPB76}.

\providecommand{\latin}[1]{#1}
\makeatletter
\providecommand{\doi}
  {\begingroup\let\do\@makeother\dospecials
  \catcode`\{=1 \catcode`\}=2 \doi@aux}
\providecommand{\doi@aux}[1]{\endgroup\texttt{#1}}
\makeatother
\providecommand*\mcitethebibliography{\thebibliography}
\csname @ifundefined\endcsname{endmcitethebibliography}
  {\let\endmcitethebibliography\endthebibliography}{}